\documentclass[prd,showpacs,amsmath,showkeys,twocolumn,floatfix]{revtex4} 
\usepackage{epsfig,dcolumn}
\usepackage{graphicx}
\usepackage[usenames]{color}
\usepackage{graphicx}
\usepackage{bm}

\def\x{{\bf x}}
\def\y{{\bf y}}

\def\k{{\bf k}}
\def\q{{\bf q}}
\def\p{{\bf p}}

\def\A{{\bf A}}
\def\B{{\bf B}}

\def\D{{\bf D}}

\def\bQ{{\bar Q}}

\def\lsim{\mathrel{\rlap{\lower4pt\hbox{\hskip1pt$\sim$}}
    \raise1pt\hbox{$<$}}}
\def\gsim{\mathrel{\rlap{\lower4pt\hbox{\hskip1pt$\sim$}}
    \raise1pt\hbox{$>$}}}
\begin{document}


\title{ Heavy quarkonium hybrids from Coulomb gauge QCD} 

\author{ Peng Guo and Adam P. Szczepaniak}
\affiliation{ Physics Department and Nuclear Theory Center \\
Indiana University, Bloomington, IN 47405, USA. }

\author{ Giuseppe Galat\`a, Andrea Vassallo and  Elena Santopinto \\
I.N.F.N. and Dipartimento di Fisica, \\
via Dodecaneso 33, Genova, I-16146, Italy. }

\date{\today}

\begin{abstract}
Using the non-relativisitc reduction of Coulomb gauge QCD we compute
spectrum of the low mass hybrid mesons
containing a heavy quark-antiquark pair. The gluon degrees of freedom
are treated in the mean field approximation calibrated to the gluelump
spectrum. We discuss the role of the non-abelian nature of the QCD
Coulomb interaction in the ordering of the spin-parity levels.
\end{abstract} 


\maketitle

\section{Introduction}

Gluons are responsible for almost the entire mass of light hadrons and
contribute significantly to hadron spin.  Gluons have been probed in
hard processes while spectroscopy of gluonic excitations has so far
been very limited to a few potential hybrid and glueball
candidates. In the near future, however experiments at JLab, PANDA and
BESIII are expected to significantly advance our knowledge of gluonic
excitations through measurements of decay channels that couple to
exotic quantum numbers carried by hybrids. PANDA and BESIII will in
particular be sensitive to charmonium hybrids.

Modern studies of gluon dominated hadron spectra have been possible
through lattice
simulations~\cite{Morningstar:1999rf,Foster:1998wu,Bali:2003jq}
while in early days of hadron phenomenology various effective models
have been developed~\cite{Horn:1977rq, Isgur:1984bm,Simonov:2000ky,Szczepaniak:1995cw,LlanesEstrada:2000hj,General:2006ed,Buisseret:2006wc,Brau:2004xw}. 
  Among these the constituent, or
quasi-particle approach have had may successes in particular in the
context of heavy quarkonia.  For gluonic excitations in presence of
heavy quarks the Born-Oppenheimer approximation can be employed to
replace the (fast) gluon field by an effective potential between the
non-relativistic quarks~\cite{Juge:2002br}. These effective
potentials originate from excited gluon configurations and can be
reliably computed using lattice gauge simulations and afterwards 
used to construct hybrid heavy quarkonia
~\cite{Juge:1999ie}. Keeping the quark-antiquark sources static and
taking their separation to zero results in a spectrum of the so
called gluelumps which describe the gluon field bound to a static
color octet source~\cite{Foster:1998wu}. It is thus natural to
expect that there should be a close relation between spectrum of
gluelumps and heavy quarkonium hybrids. This is because in the
latter the average spacial separation between heavy quarks is small,
of the order of $\langle r \rangle \sim O(1/(\alpha_s m))$ where,
$m$ is the heavy quark mass and $\alpha_s$ is the running coupling
evaluated at distance scale $O(\langle r \rangle)$.
    
Recent lattice simulations of the charmonium spectrum in quenched
QCD beyond the Born-Oppenheimer approximation also lead to more
states then predicted by the quarkonium potential models
potentially indicating presence of gluonic
excitations~\cite{Dudek:2007wv}.  In the quenched approximation
and with non-relativistic quarks the connection between heavy
quarkonium spectrum and the potential model can be made
precise. In the context of the Wilson loop formalism this has
been done in
~\cite{Eichten:1980mw,Simonov:1988mj,Barchielli:1988zp} where
potential matrix elements between heavy quarkonium states were
related to correlation functions containing QCD gluon field
operators. In ~\cite{Feinberg:1977rc} the Foldy-Wouthousen
formalism was used instead and the effective non-relativistic
QCD Hamiltonian in the Coulomb gauge was derived. In this
framework all operators conserve the heavy quark quark number and
depend on the gluon field. The effective $Q{\bar Q}$ potential
matrix elements in general depend on the distribution of the
gluon field in the system.  In general, gluon distribution in the
vacuum and in a state containing a $Q{\bar Q}$ pair are
different. In~\cite{Greensite:2003xf} it was found that at large
separations the Coulomb energy exceeds that of true static
quarkonium state.  This makes possible to interpret the former
as a variational approximation to the true $Q{\bar Q}$ QCD
state. This difference has simple interpretation in the Coulomb
gauge~\cite{Szczepaniak:2006nx,Szczepaniak:2005xi}. The Coulomb
energy corresponds to the energy of an approximate, variational
state with the $Q{\bar Q}$ pair added to the QCD
vacuum. Gluon-number changing interactions in the Coulomb
potential couple the $Q{\bar Q}$ source to an arbitrary number of
transverse gluons which eventually produces the string-like QCD
$Q{\bar Q}$ eigenstate whose energy is that of the Wilson
loop. Extrapolating to small $Q{\bar Q}$ separations, however, a
reasonable initial approximation would be to assume that the
gluon field is not significantly disturbed by the heavy quark
sources and to build the spectrum of gluon excitation using an
effective gluon Fock space that is independent from the
sources. Such a spectrum was constructed in
~\cite{Szczepaniak:2001rg, Szczepaniak:2003ve,Feuchter:2004mk,
  Reinhardt:2004mm,
  Schleifenbaum:2006bq,Epple:2006hv,Epple:2007ut} using a
variational mean field approximation and describes effective
massive gluons propagating in a nonperturbative vacuum. We will
use this Fock space here and combine it with non-relativistic
quark an antiquark to compute spectrum of heavy hybrids.

In ~\cite{Guo:2007sm} we followed the same approach to compute  
spectrum of gluelumps. By using a single framework offered by the
non-relativistic QCD, we will be able to explore the connection
between gluelump and hybrid spectra.  In this paper we consider the
leading contribution to the heavy quarkonium hybrid energies {\it
  i.e.} excluding spin dependent-effects. In the following Section
we discuss derivation and treatment of the Foldy-Wouthousen Coulomb
gauge (FWCG) Hamiltonian and the mean field approximation to the
QCD vacuum. In Section ~\ref{g-h} we summarize calculation of
spectrum of ordinary (non-hybrid) quarkonia which is used later to
benchmark the hybrid spectrum which is studied in
Section~\ref{hybrids}.  Summary and outlook are given in
Section~\ref{outlook}.

\section{ Foldy-Wouthousen Coulomb gauge Hamiltonian} 
\label{FWCG}

The non-relativistic QCD Hamiltonian, $H_{FWCG}$ describing interaction between heavy quarks and gluons  can be constructed from the full QCD Hamiltonian in the Coulomb gauge by employing Foldy-Wouthuysen transformation~\cite{Feinberg:1977rc}
\begin{equation}
H \to H_{FWCG} = T H T^{-1}.
\end{equation}
 Here  $T= \cdots e^{iS_2} e^{iS_1} e^{iS_0}$ and the hermitian generators $S_i$ are of the order $(\langle p \rangle/m)^i$ respectively , with $m$ standing for  mass of the heavy quarks and $\langle p \rangle$ being the average three-momentum of the heavy quarks. The FWCG Hamiltonian is thus an effective Hamiltonian valid for momenta smaller  than the heavy  quark mass.  The operators $S_i$  eliminate couplings between upper and lower components of  Dirac spinors to any given order in the $1/m$ expansion and lead to a Hamiltonian that conserves number of (heavy) quarks. It 
   has the form of  a series expansion in inverse powers of the heavy quark mass, 
 \begin{equation} 
H_{FWCG} =   H_{-1} + H_0 +  H_1 + H_2 + \cdots 
\end{equation} 
Each term, $H_i$ being of order $m^i$  depends on the transverse gluon  and the  non-relativistic quark and antiquark fields. The transverse gluon degrees of freedom are given by the vector  potential 
 $\A^a(\x)$ and the conjugated momentum $\bm{\Pi}^a(\x)$  ($\bm{\nabla}\cdot\A^a = \bm{\nabla}\cdot\bm{\Pi}^a = 0$) satisfying 
\begin{equation}
[ \A^a(x), \bm{\Pi}^b(\y) ] = \bm{\delta}_T(\x - \y)\delta^{ab} 
\end{equation}
  where  $\bm{\delta}_T(\x-\y) \equiv [{\bf I}  - \bm{\nabla} \bm{\nabla}/\bm{\nabla}^2] \delta^3(\x-\y)$). 
 The canonical momentum $\bm{\Pi}$ is the  negative of the transverse component of the chromo-electric field. 
 The quark and antiquark  degrees of freedom can be written directly in terms of  quark creation and annihilation $Q^{\dag},Q$) and antiquark creation and annihilation, (${\bar Q}, {\bar Q}^{\dag}$ 
   operators which in terms of the original Dirac fields are given by, 
 \begin{eqnarray}
& &  Q_\lambda(\x) \equiv  \left[{{1 + \beta}\over 2}  \psi(\x)\right]_\lambda,    \nonumber \\
& & [{\bar Q}(\x) (-i\sigma_2) ]_\lambda \equiv  \left[\psi^{\dag}(\x){ {1- \beta}\over 2}\right]_{\lambda+2}. 
 \end{eqnarray} 
 Here $\lambda=1,2$ denotes the $z$-component of quark spin and the Pauli matrix $\sigma_2$  is introduced so that  $Q_\lambda$ and ${\bar Q}_\lambda$ belong the the same $SU(2)$ representation.  The mass term $H_{-1}$ is given by (summation over spin indices is implicit) 
 \begin{equation}
 H_{-1} =  m \int d\x \left[ Q^{\dag}(\x) Q(\x) + \bQ^{\dag}(\x) \bQ(\x) \right].
 \end{equation} 
 The $H_0$ term represents the Yang-Mills Hamiltonian  coupled to static quark and antiquark sources, 
   \begin{eqnarray}
 H_0 & = &  {1\over 2} \int d\x \left[ {\cal J}^{-1} \bm{\Pi}^a(\x) {\cal J}  \bm{\Pi}^a(\x) + \B^a(\x) \cdot \B^a(\x)  \right] 
 \nonumber \\ 
  & + & {1\over 2} \int d\x d\y  {\cal J}^{-1} \rho^a(\x) K(\x,a;\y,b) {\cal J} \rho^b(\y). \label{YM}
  \end{eqnarray} 
 Here ${\cal J} = \mbox{Det}\bm\left[-{\nabla}\cdot  {\cal D}\right]$ is the determinant of the Faddeev-Popov operator; ${\cal D} = {\cal D}_{ab} = \delta_{ab} \bm{\nabla}  + g f_{acb} {\bf A}^c$  is the covariant  derivative in the adjoint representation, and $\B$ is the chromo-magnetic field, $\B^a(\x) =  \bm{\nabla} \times \A^a(\x) + (g/2) f_{abc} \A^b(\x) \times \A^c(\x)$. The last term in Eq.~(\ref{YM}) represent the non-abelian Coulomb gauge interaction between color charge densities, $\rho^a(\x)  = Q^{\dag}(\x) T^a Q(\x) - \bQ^{\dag}(\x) T^{*a} \bQ(\x) + f_{abc}  \bf{A}^b(\x) \bm{\Pi}^c(\x) $, 
and the Coulomb kernel is given by
\begin{equation}
K(\x,a;\y,b) = \left[ {g \over {\bm{\nabla} \cdot {\cal D} }} (-\bm{\nabla}^2) {g \over { \bm{\nabla} \cdot {\cal D}}}  \right] _{(\x,a;\y,b)}. \label{kernel}
\end{equation} 
The $O(1/m)$ terms are given by 
\begin{widetext} 
\begin{equation} 
H_1  =  {1\over {2m}} \int d\x  \left[ Q^{\dag}(\x) \D^2(\x) Q(\x)  + (Q \to {\bar Q}, T \to T^*)  \right] 
  - {1\over {2m}}  \int d\x  \left[ Q^{\dag}(\x) g\bm{\sigma}\cdot \B^a(\x) T^a Q(\x)  - (Q \to {\bar Q}, T \to T^*) \right]\nonumber \\ \label{h1} 
 \end{equation} 
 \end{widetext}
with $\D$ being the covariant derivative in the fundamental representation, 
$\D = \D_{ij}(\x) = -i\bm{\nabla}\delta_{ij} - gT^a_{ij} \A^a(\x)$. 
  \begin{figure}
\includegraphics[width=2in,angle=0]{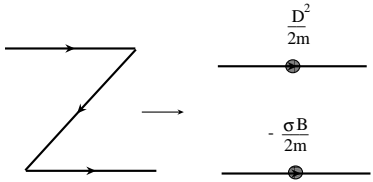}
\caption{\label{z} Non-relativisitc reduction of heavy quark propagator  (left side) leads to a kinetic energy term and  hyperfine interaction between quark  (chromo)magnetic  moment and the (chromo)magnetic field. } 
 \end{figure} 
The $O(1/m)$  Hamiltonian arises from the  $Z$-graph  (see Fig.~\ref{z}) and is generated by the first order FW transformation with $S_1 = (-i/2m) \int d\x \psi^\dag(\x) \beta\bm{\alpha} \cdot \D \psi(\x)$. The FW transformation effectively  eliminates the quark propagator in Fig.~\ref{z} and shrinks the relative distance between the gluon interaction points to zero. This requires that the momentum of the quark propagator is smaller then $m$  thus, as already mentioned, it is necessary that that quark momenta are smaller then the quark mass ( 
this is expected to be the case for low mass heavy quarkonium bound  states
).  The first term in $H_1$ contains quark and anti-quark kinetic energies and $O(1/m)$ local, spin-independent interaction between (anti)quark  and gluon. The second terms  describes $O(1/m)$ spin-dependent local interaction between (anti)quark  and gluon. In the leading order we drop the $O(1/m)$ and higher interactions terms and keep only kinetic energies of the quarks. In Section ~\ref{hybrids} we comment on the results of next to leading order calculation, were we add the remaining $O(1/m)$ terms.  Finally the second term contains $O(1/m)$ interaction that mixes pure $Q{\bar Q}$ states with $Q{\bar Q}g$, hybrid components. These generate $O(1/m^2)$ spin-depdnent splitting  which we ignore all together. The final form of $H_1$  which is diagonal in the hybrid basis, and is  leading order in $O(1/m)$  
    is thus given by 
\begin{widetext} 
\begin{eqnarray} 
H_1 & = &  -\frac{1}{2m} \int d\x \left[  Q^{\dag}(\x)\bm{\nabla}^2 Q(\x) + {\bar Q}^{\dag}(\x)
\bm{\nabla}^2 {\bar Q}(\x)  \right]
 +  {{g^2} \over {2m}} \int d\x \left[Q^{\dag}(\x) T^a T^b \A^a(\x) \cdot \A^b(\x)  Q(\x)  + (Q \to {\bar Q}, T \to T^*)  \right] \nonumber \\
 & - &  {{g^2} \over {4m}}  \int d\x  \left[ Q^{\dag}(\x)   T^a f_{abc} \bm{\sigma} \cdot 
 \A^b(\x) \times \A^c(\x) Q(\x)  - (Q \to {\bar Q}, T \to T^*) \right]\nonumber \\ \label{h1last} 
 \end{eqnarray}
\end{widetext} 
with only the first term used in what we define as the leading order calculation. 

 \subsection{Gluonic degrees of freedom in the mean field} 

 In the  Shr\"odinger representation the lowest eigenvalue of the QCD Hamiltonian $H_{QCD}[A] \Psi_n[A] = E_n \Psi[A]$ , $\Psi_0[A]$ represents  transverse gluon field distribution in the vacuum. 
In the Coulomb gauge Fock space construction of the excited states 
(mesons, baryons, hybrids, {\it etc \ldots})   
 is motivated by the Gribov-Zwaniziger  confinement scenario. The vacuum state is expected to be dominated by field configurations near boundary  of the Gribov horizon, and gluon excitations which build the hadron spectrum 
  correspond to massive quasi-particles of the harmonic expansion of  the the wave functional  near the Gribov horizon. In practice, in a mean field description~\cite{Szczepaniak:2001rg,Szczepaniak:2003ve,Feuchter:2004mk,Reinhardt:2004mm,Schleifenbaum:2006bq,Epple:2006hv,Epple:2007ut} the dominance of the field configurations near the Gribov horizon is obtained by choosing the renormalized coupling $g(\Lambda)$ to be such that the inverse of the Faddeev-Popov develops a pole near the zero mode. 
Consider a gaussian ansatz for the vacuum wave functional,  
\begin{equation} 
\Psi_0[A] =  \langle \A|0\rangle \equiv \exp\left( - {1\over 2}  \int {{d\k} \over {(2\pi)^3}} \A^a(\k) \omega(k) \A^a(\k)  \right), \label{vac} 
 \end{equation}
 \begin{figure}[ptbh]
\begin{center}
\includegraphics[width=0.45\textwidth]{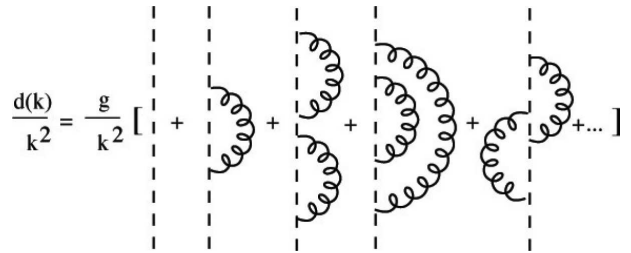}
\end{center}
\caption{Diagrammatic representation of the expansion of the functional integral for the ghost propagator {\it c.f.} Eq.~(\ref{d})} 
\label{rainbow}
\end{figure}
 in the rainbow ladder approximation, show in Fig.~\ref{rainbow} the expectation value of the inverse Faddeev-Popov operator 
 \begin{equation} 
 \int d\x e^{i\k\cdot\x} \langle \Psi_0|{g \over {-\bm{\nabla}\cdot{\bf D}}}| \Psi_0\rangle \equiv {{d(k)} \over {k^2}}, \label{d}   
 \end{equation}
satisfies, ($\hat\k \equiv \k/k$), 
\begin{equation} 
{1\over {d(k)}} = {1\over {g(\Lambda)}} - {N_C \over 2} \int^\Lambda {{d\p} \over {(2\pi)^3}} 
(1 - \hat\k\cdot\hat\p) {{d(|\p-\k|) } \over {\omega(p) (\k - \p)^2}}. \label{fp} 
\end{equation} 
The numerical solution to this Dyson equation has been extensively studied in \cite{Szczepaniak:2001rg,Szczepaniak:2003ve,Feuchter:2004mk,Reinhardt:2004mm,Schleifenbaum:2006bq,Epple:2006hv,Epple:2007ut}. The main features of its solutions remain unchanged if one uses the angular approximation in evaluating the momentum integral
$ |\p - \k| \to p \theta( p - k) + k \theta( k - p)$  and use the approximation 
$\omega(k) \to m_g (=\mbox{constant})$ for $k  \to 0$ and $\omega(k) \to k$ for $k > m_g$~\cite{Szczepaniak:2001rg}. 
This approximation  enables to obtain an  analytical solution to Eq.~(\ref{fp}) which is given by 
\begin{equation} 
d(k) =  \left\{ \begin{array}{c} 
{{g(\Lambda)} \over {\left[1 + \frac{\beta_L}{(4\pi)^2}  g^2(\Lambda) 
\left(  \frac{k}{m_g} - \frac{\Lambda}{m_g} \right) \right]^{1/2}}} ,  k < m_g   \\ 
  {{g(\Lambda) } \over {\left[1 + \frac{\beta_H}{(4\pi)^2}  g^2(\Lambda) 
  \log\left( \frac{k^2}{m^2_g}\right)\right]^{1/2}}}, k > m_g 
\end{array}  \right.
\end{equation} 
with $\beta_H = 8N_C/3  $ and $\beta_L = 5\beta_L/3 $. It follows that $d(k)$ can be made renormalization scale  invariant by choosing $g(\Lambda)$ to depend on $\Lambda$  in the same way $d(k)$ depends on $k$. This also allows for interpreting $d(k)$ as the running coupling.  For $g(\Lambda) \le g_C \equiv 4\pi (m_g/ \beta_L \Lambda )^{1/2}$ the Landau pole is avoided and, as $g(\Lambda) \to g_C$ from below, the functional integral of the gluon field in Eq.~(\ref{d})  becomes dominated by the configurations near the Gribov horizon. The parameter $\omega$ which fixes the vacuum wave functional can be constrained by minimizing the energy density $\delta \langle \Psi_0|H|\Psi_0\rangle = 0$. \\ 

The resulting gap equation involves  expectation value of the Faddeev-Popov operator whose computation, however is not free of  from ambiguities. These ambiguities arise of UV subtraction which can be absorbed into definition of $\omega$~\cite{Reinhardt:2004mm,Epple:2007ut}.  Therefore we choose a more phenomenological approach. 
 Following our studies of the gluelump  spectrum~\cite{Guo:2007sm}  we will  treat $\omega$ as a parameter of the model to be constrained by the physical spectrum rather then by the properties of the vacuum {\it i.e.} the gap equation. 
Since the gluon 2-point function is given in terms of $\omega$, by 
\begin{equation} 
\langle \Psi_0 | \A^a(\k) \A^b(\p) \rangle = (2\pi)^3\delta(\k + \p) {{\delta_{ab}} \over {2\omega(k)}} 
\end{equation} 
it is clear that the wave functional in Eq.~(\ref{vac}) leads to a gap in single gluon spectrum. The energy of the single gluon state defined as 
\begin{equation} 
|1g\rangle = |\k,\lambda,a\rangle = a^{\dag}(\k,\lambda,a) |\Psi_0\rangle
\end{equation}
where the single gluon creation (annihilation) operators $a^{\dag}$,($a$) are defined by ($\bm{\epsilon}$ are helicity vectors) 
\begin{widetext} 
\begin{eqnarray} 
\A^a(\x) = \int {{d\k}\over {(2\pi)^3} }  {1\over {\sqrt{2\omega(k)}}} 
\left[ \bm{\epsilon}(\k,\lambda) a(\k,\lambda,a) + \bm{\epsilon}^{\dag}(-\k,\lambda) a^{\dag}(-\k,\lambda)\right] e^{i\k\cdot\x} \nonumber \\
\bm{\Pi}^a(\x) = -i \int {{d\k}\over {(2\pi)^3} }  \sqrt{{ \omega(k)} \over 2} 
\left[ \bm{\epsilon}(\k,\lambda) a(\k,\lambda,a) - \bm{\epsilon}^{\dag}(-\k,\lambda) a^{\dag}(-\k,\lambda)\right] e^{i\k\cdot\x} \nonumber \\
  \end{eqnarray} 
  \end{widetext}  
  is given by ~\cite{Szczepaniak:2001rg,Szczepaniak:2003ve}
  \begin{equation} 
\frac{ \Sigma_g(k)}{\omega(k)} = 1  -   \frac{N_c}{2 } (2 \pi) \int \frac{ d\k' }{(2 \pi)^3}
 \widetilde{V}_{CL} (|\k - \k'|)  \frac{ 1+(\hat \k\cdot \hat \k')^2}{2\omega(k')}. \label{eg} 
\end{equation} 
Here $\widetilde{V}_{CL}$ is the negative of the momentum space expectation value of the Coulomb potential 
\begin{equation} 
\widetilde{V}_{CL}(k) = \int d\x e^{i\k\cdot\x} \langle \Psi_0 | {g \over {\bm{\nabla}\cdot{\bf D}}} (\bm{\nabla}^2) {g \over {\bm{\nabla}\cdot{\bf D}}}  | \Psi_0 \rangle  \label{vcl} 
  \end{equation} 
  which is well approximated by a combination of Coulomb and linear potential ~\cite{Szczepaniak:2001rg}. The long range part of $V_{CL}$ can potentially lead to and IR divergence in the self energy ( integral in Eq.~(\ref{eg}))  from the integration region $\k' \sim \k$. This divergence is  however canceled in the full bound state equation by similar  divergencies  arising from quark self energies and quark-antiquark interaction. The quadratic  UV divergence from $\k' \to \infty$ in the self energy integrals is canceled by a gluon mass counter-term ~\cite{Szczepaniak:2001rg,Szczepaniak:2003ve}.  In the basis of massive gluons the Foldy-Wouthousen  Hamiltonian is schematically given by 
\begin{equation} 
H_{FWCG} = \sum_{\k,\lambda,a}  \Sigma_g(k) a^{\dag}(\k,\lambda,a)a(\k,\lambda,a) + V(a^\dag,a) 
\end{equation} 
where $V$ represents residual interactions between gluons and heavy quarks. The gluon mass gap implied by 
$\Sigma_g(k) \ne k$ suggests that a hybrid state in which quarks and gluons have residual 3-momenta of the order of $\Lambda_{QCD}$ should be well approximated by a state containing a pair of  non-relativisitc  quark and antiquark and a single quasi-gluon. In the infinite heavy quark mass limit, separation between the quarks becomes a good quantum number and energy spectrum of such {\it static} hybrids is determined by the energy of the single quasi-gluon in presence of color octet $Q{\bar Q}$ source. As separation between quarks is taken to zero spectrum of static hybrids becomes rotationally invariant and such states are being referred to as gluelumps. 
 Lattice simulations show that  the lowest glue-lump state has spin J, parity P, and charge  conjugation  C, $J^{PC}=1^{+-}$ and the first excited state, $J^{PC} = 1^{--}$. This unusual parity inversion between vector and  pseudo-vector  states has been explained using the Coulomb gauge approach as due to the three-body force involving all three particles quark, antiquark and gluon originating from the non-abelian Coulomb interaction in Eq.~(\ref{YM}). A good agreement between our Coulomb gauge approach and lattice gluelump and static hybrid spectra gives us confidence that  the approach may also be adequate for computation of the spectrum of hybrids with dynamical heavy quarks.

\section{ Ordinary quarkonia} 
\label{g-h}
Before discussing gluonic excitations we shall fix the parameters of the $Q{\bar Q}$ sector by comparing the 
$H_{FWCG}$ Hamiltonian spectrum with that of ordinary charmonia. 

\subsection{Basis and Hamiltonian matrix elements} 
The Coulomb gauge Foldy-Wouthuysen Hamiltonian should reproduce the known spectrum of ordinary quarkonia. 
 Since low-lying quarkonia are expected to be dominated by the $Q{\bar Q}$  component  we may represent $N$-th  quarkonium state  of spin and $z$-component  $J,M$ parity $P$ and charge conjugation $C$, as 
\begin{equation}
|JMPCN \rangle = \sum_{\alpha \equiv (S_q,L_q)} \int \frac{q^{2}dq }{(2
\pi)^{3}} \Psi^{N}_{\alpha} (q) |JMPC;\alpha;q \rangle,
\end{equation}
with the $Q{\bar Q}$ state given by 
\begin{eqnarray}
|JMPC;\alpha;q \rangle & =& \sum_{m_{1},m_{2}} \int  d \hat{\mathbf{q}}
 \chi^{JMPC}_{m_{1},m_{2}}   (\hat{\mathbf{q}},\alpha) \nonumber \\
 & \times & Q^{\dag}_{\mathbf{q},m_{1},i_{1}}
\frac{ \delta_{i_{1},i_{2}} }{\sqrt{N_{c}} }{\bar Q}^{\dag}_{-\mathbf{q},m_{2},i_{2}}
 |0\rangle,
\end{eqnarray}
and where the spin-orbial wave function $\chi$ describes coupling of total quark spin $S_q$ to relative orbital angular momentum $L_q$ to total spin of  the quarkonium $J$, ${\bf L}_q + {\bf S}_q = {\bf J}$, 
\begin{widetext} 
\begin{equation} 
\chi^{JMPC}_{m_{1},m_{2}} (\hat{\mathbf{q}},\alpha)   = 
  \frac{1}{2} [1+C(-1)^{L_q+S_q}] \frac{1}{2} [1 + P(-1)^{L_q+1}]  
 \sum_{M_{S},M_{L}} \langle \frac{1}{2} 
m_{1};\frac{1}{2} m_{2}|S_q M_{S}\rangle 
\langle S_q M_{S};L_q M_{L}|J M \rangle   
Y_{L_q M_{L}}(\mathbf{q}) .
\end{equation}
\end{widetext} 
The Shr\"odinger equation for the masses of ordinary quarkonia is given by 
\begin{eqnarray} 
 M_N \Psi_\alpha^N(q)  & =  & \left[ 2m + \frac{q^2}{m} +  \Sigma_q\right] \Psi_\alpha^N(q)  \nonumber \\
  & +  & C_F \sum_{\alpha'} \int \frac{q'^2 dq'}{(2\pi)^3} V_{Q\bar Q}(q,\alpha;q'\alpha') \Psi_{\alpha'}^N(q')  \nonumber \\ \label{sqq} 
\end{eqnarray} 
The first  term represents quark and antiquark kinetic energies and their self-energies, 
\begin{equation} 
\Sigma_q = -C_F V_L(0) = -C_F \int \frac{d\q}{(2\pi)^3} \widetilde{V}_{L}(q) \label{sq} 
\end{equation} 
In position space the expectation value of the Coulomb potential given in Eq.~(\ref{vcl}) leads to a short range part 
$V_{CL}(R) \to V_C(R) \propto -1/R$ for $R \to 0$ and long range part, $V_{CL}(R) \to V_L(R) \propto R$ for $R \to \infty$ 
The contribution from the short range part is absorbed in the definition of the quark mass. The long range part leads to an IR divergence which  in momentum space and is represented by the constant $V_L(0)$ in Eq.~(\ref{sq}). This constant, and therefore sensitivity to the $R\to \infty$ limit of the position space Coulomb interaction is removed by the long-range part of the quark-antiaquark potential, 
\begin{equation}
V_{Q\bar Q} = \delta_{\alpha'\alpha} \int d^2\hat\q' P_L(\hat\q'\cdot\hat \q) \widetilde{V}_{CL}(|\q'-\q|) 
\end{equation} 
as it should for color singlet bound states. 

We use the  quarkonium  spectrum to determine the renormalized quark mass. 
From  experimental data  we find for the  the  $S$-wave  ($L_q=0$) spin-averaged  
masses  
\begin{equation} 
\bar{M}^{S}=\frac{1}{4}[M_{ 0^{-+}  }+3 M_{1^{--} }] 
\end{equation} 
$\bar{M}^{S}_{c \bar{c}} =3.068 $ Gev for charmonium and $\bar{M}^{S}_{b \bar{b}}=9.46$ Gev for bottomonium, respectively. Comparing with results of numerical solution to Eq.~(\ref{sqq}) we extract  $m_{c}=1.16\mbox{ GeV}$ and $m_{b}=4.58\mbox{ GeV}$, respectively. In Fig.~\ref{cc} we compare the predicted masses of the first radial  excitation  of the $S$-waves and the ground state (spin-averaged) $P$-waves with experimental data. The observed agreement  is a reflection of the Coulomb gauge kernel correctly reproducing the "Coulomb + Linear" potential  in the  region  of position space covered by the bound quarkonium wave functions. 

\section{Spectrum of Gluonic Excitations }
\label{hybrids} 

As discussed above, gluelumps provide a good benchmark for the spectrum of gluonic excitations. Upon neglecting  the quark motion ({\it i.e} setting $H_1=0$)  (anti)quark positions become good quantum numbers, and in particular the  spectrum of  $Q{\bar Q}$ states with vanishing relative separation between quark sources  is given by the energy of gluon cloud. In our approach the low lying spectrum is obtained by binding  a single quasi gluon with kinetic energy given by Eq.~(\ref{eg}) to the $Q{\bar Q}$ source through the Coulomb kernel given in Eq.~(\ref{YM}). More details are given in Ref.~\cite{Szczepaniak:2006nx,Szczepaniak:2005xi,Guo:2007sm}. The spin of the gluelump is then identified with the total angular momentum $J_g$ of the quasi-gluon, orbiting the static $Q{\bar Q}$ source. Once the sources are allowed to move  the total spin of the resulting hybrid meson is given by the sum of angular momenta of all constituents. Since heavy quarks are expected to move non-relativisitcaly with average separation of the order of $O(1/\alpha_s m)$ it is reasonable to expect there will be not much distortion of the gluon wave function of the gluelump due to quark motion. Finely, since we are ignoring quark-spin dependent interactions it will be useful to couple quark spin at the very last to other 
angular momenta before producing a state of good total spin $J$.

\subsection{ Basis and Hamiltonian matrix elements } 

From the above discussion it follows that the  optimal basis of $Q{\bar Q}g$ states  is such in which the $Q{\bar Q}$ relative  angular momentum $L_q$ is first coupled to total gluon spin $J_g$. The resulting angular momentum $j$  is then coupled  to the total quark-antiquark spin $S_q$  to the give the total spin of the hybrid state $J$.  In this coupling scheme it is straightforward to take the static quark limit and thus  compare hybrid and gluelump spectra. Specifically, the $N$-th hybrid state with total spin and its $z$-axis projection,$J,M$, parity $P$, charge conjugation $C$ is given by 

\begin{eqnarray}
|JMPCN  \rangle  & =  &   \sum_{\alpha \equiv (\sigma,S_q,L_q,J_q)} \int \frac{k^{2}dk
}{(2 \pi)^{3}} \frac{q^{2}dq }{(2 \pi)^{3}} \Psi^{N}_{\alpha} (k,q) \nonumber \\
  &\times  &  |JMPC;\alpha;k,q \rangle .
\end{eqnarray}
Here the quark-antiquark-gluon, $Q{\bar Q}g$ state is given by 
\begin{eqnarray}
&&|JMPC; \alpha;k,q \rangle   =    \sum_{m_{1},m_{2},\sigma} \int d
\hat{\mathbf{k}} d \hat{\mathbf{q}}
\chi^{JMPC}_{m_{1},m_{2},\sigma}
(\hat{\mathbf{k}},\hat{\mathbf{q}},\alpha) \nonumber \\
 & \times &   Q^{\dag}_{\frac{\mathbf{k}}{2}+\mathbf{q},m_{1},i_{1}}
 \frac{T^{a}_{i_{1},i_{2}}}{\sqrt{C_{F} N_{c}} }
{\bar Q}^{\dag}_{\frac{\mathbf{k}}{2}-\mathbf{q},m_{2},i_{2}}
a^{\dag}_{-\mathbf{k},\sigma,a} |0\rangle .  \nonumber \\
\end{eqnarray}
The spin-orbital wave function $\chi$ describes the $({\bf L_q}  + {\bf  J}_q ) +  {\bf S}_q $ coupling discussed above , and $\sigma = \pm 1$ represents gluon  helicity 
\begin{eqnarray}
&&\chi^{JMPC}_{m_{1},m_{2},\sigma}
(\hat{\mathbf{k}},\hat{\mathbf{q}},\alpha)  = 
\sqrt{\frac{2J_g+1}{4 \pi}}  \frac{1}{2} [1+C(-1)^{L_q+S_q+1}]  \nonumber \\
& \times & \frac{1}{\sqrt{2}} (-1)^{ J_g} \sum_{M_{S},M_{L},m} \langle \frac{1}{2}
m_{1};\frac{1}{2} m_{2}|S_q M_{S} \rangle  \nonumber \\
& \times & \langle J_g M_g; L_q M_L | j m \rangle \langle  j m ;S_q M_{S}|J M \rangle  Y_{L_q M_{L}}(\mathbf{q})  \nonumber \\
&\times& 
 D^{\ast J_g}_{M_g,
-\sigma}(\hat{\mathbf{k}}) [ \delta_{\sigma,1} + P(-1)^{J_g+L_q+1}
\delta_{\sigma,-1}] .  \label{hs} 
\end{eqnarray}

Finally, $\q$ is the relative momentum between the quark-antiquark and $\k$ is the momentum of the gluon in the hybrid center of mass frame. The spin-orbtal wave function is kinematical, {\it i.e.} determined by the rotational symmetry of $H_{FWCG}$ Hamiltonian, while the radial  wave function $\Psi^N_\alpha(k,q)$ will be determined by  diagonalizing  the Schr\"odinger equation for the hybrid bound state. Parity and charge conjugation are also kinematical and given by 
\begin{eqnarray} 
P & = &  \xi (-1)^{J_g + L_q + 1},   \nonumber \\
C & = & (-1)^{L_q + S_q +1}, 
\end{eqnarray} 
respectively. Here $\xi = +1$ for  TM (natural parity) and $\xi =-1$ for TE (  unnatural  parity ) gluon state which 
 correspond to a $|\sigma = +1\rangle + \xi |\sigma = -1\rangle$ combination of gluon  helicity states. 
We note that, as expected,  both $P$ and $C$ are a product of $Q{\bar Q}$ and gluelump  parity  and charge conjugation given by 

\begin{eqnarray}
P_q  & = &  (-1)^{L_q + 1}, \; P_g = \xi (-1)^{J_g}, \nonumber \\
C_q & =  & (-1)^{L_q + S_q }, \; C_g = -1. 
\end{eqnarray} 
The Schr\"odinger equation for the radial wave function has the form of 
\begin{widetext} 
\begin{eqnarray} 
\left[2m  + \frac{q^2}{m} + \frac{k^2}{4m} + \Sigma_g(k)  + \Sigma_q \right] \Psi^N_{\alpha}(k,q)  & - &  
\frac{1}{2N_C} \sum_{\alpha'} \int \frac{q'^2dq'}{(2\pi)^3} V_{Q\bar Q}(q,\alpha;q',\alpha') \Psi^N_{\alpha'}(k,q') \nonumber \\
&  + & 
\sum_{\alpha'} \int \frac{k'^2dk'}{(2\pi)^3}   \frac{q'^2dq'}{(2\pi)^3} V_{Q{\bar Q}g}(k,q,\alpha;k',q',\alpha') \Psi^N_{\alpha'}(k',q')  = 
M_N  \Psi^N_{\alpha}(k,q).  \nonumber \\ \label{seq}
\end{eqnarray} 
\end{widetext} 
In leading order the fist term represents kinetic energies and self-energies  of the gluon,  given by  Eq.~(\ref{eg}), and of quarks given by Eq.~(\ref{sq}).  As in the case of Eq.~(\ref{sqq}) the IR divergencies in the self energies are canceled by the  
 potential matrix elements.  The first integral in Eq.~(\ref{seq}) 
 is the $Q{\bar Q}$, interaction. It is independent of the gluon degrees of freedom and repulsive in for the $Q{\bar Q}$ pair in the color octet.  Finally, the last term on the left hand side of Eq.~(\ref{seq}) represents collectively the two body  attractive interactions between the quark and the gluon and the antiquark and the gluon and the irreducible three-body interaction linking all three constituents -- quark, antiquark and gluon, 
 \begin{equation} 
V_{Q{\bar Q}g} = V^2_{Q{\bar Q}g} + V^3_{Q{\bar Q}g}. \label{vq} 
\end{equation}
The two-body interaction is given by 
\begin{widetext} 
\begin{eqnarray}
V^2_{Q \bar{Q} g}(k,q,\alpha;k',q',\alpha)  &=&    \frac{N_c}{4 } \left( \sqrt{\frac{\omega(k)}{
\omega(k')}}  + \sqrt{\frac{\omega(k')}{ \omega(k)}} \right)    \sum_{ m_{1},m_{2},\sigma, \sigma'}
 \int d \hat{\mathbf{k}} d
\hat{\mathbf{q}}  d \hat{\mathbf{k}}' d \hat{\mathbf{q}}'
\chi^{\ast JMPC}_{m_{1},m_{2},\sigma}
(\hat{\mathbf{k}},\hat{\mathbf{q}},\alpha)
\chi^{J'M'P'C'}_{m_{1},m_{2},\sigma'}
(\hat{\mathbf{k}'},\hat{\mathbf{q}}',\alpha')  \nonumber \\
&\times& 
\left[ \delta(\q +\k/2 - \q' - \k'/2)+\delta(-\q  +\k/2+ \q' -\k'/2) \right] 
  \widetilde{V}_{CL}(\mathbf{k} -\mathbf{k}')  \epsilon_{-\mathbf{k}',\sigma'} \cdot
\epsilon^*_{-\mathbf{k},\sigma} .
\end{eqnarray}
\end{widetext} 
The three body interaction  emerges from the non-abelian Coulomb kernel in Eq.~(\ref{kernel}). 
The  general expression for the three body interaction,$V^3_{Q{\bar Q}g}$  
 is quite complicated, and we discuss special cases in the Appendix. The  matrix elements of next to leading order interactions are also given in the Appendix. 

\subsection{Hybrid vs gluelump spectra} 
The particular choice of the coupling of spin and orbital degrees of freedom in Eq.~(\ref{hs}) enables to make a direct contact with the gluelump spectrum. In that case the orbital part of the  $Q{\bar Q}$ wave function is in the $S$-wave 
 and the  radial wave function  is proportional to a $\delta$-function in coordinate space. We take the radial part as a normalized  gaussian, 
 \begin{equation} 
 \Psi_Q^S(q) = \frac{2}{\pi^{\frac{1}{4}}} \left( \frac{2\pi}{\alpha_Q} \right)^{\frac{3}{2}} e^{-\frac{q^2}{2\alpha_Q^2}}, \;\;\ \int\frac{q^2 dq}{(2\pi)^3} |\Psi_Q^S(q)|^2 = 1.  \label{s-wave} 
\end{equation}
 The leading order Hamiltonian truncated to the $S$ -wave quark-antiquark orbital is given in the Appendix. 
 It is straightforward to verify that removing the quark mass term and taking  the limit $\alpha_Q \to \infty$  ( after  $m\to \infty$  )  in Eq.~(\ref{seq}),  which corresponds to  setting the relative quark separation to zero,  
 the gluelump Schr\"odindger equation from ~\cite{Guo:2007sm}  is reproduced.  

With dynamical quarks the quark-gluon interaction mixes the quark orbital angular momentum with gluon spin while in leading order it preserves quark spin. This is represented  by a mixing between $\alpha$ and $\alpha'$ indices. To leading order, however, quark spin is conserved, $S_q = S'_q$ and then  parity and charge conservation imply no mixing between TE and TM gluons.   As long as the quark motion does not distort  much the gluon distribution of a gluelump one would expect  mixing between various $(L_q,J_g)$ states to be small.  Spin splitting  is also expected  to be small,  of the order of $O(1/m)$. In other words spectrum of 
  low hybrids are expected to be approximately classified by the product of gluelump and $Q{\bar Q}$ quantum numbers. 
   In the following we will compute the hybrid spectrum under this approximation. 
   Do not see much point in going beyond this approximation. This  because  we expect  uncertainty in  the spectrum associated with our mean field   treatment of gluonic degrees of freedom  to be of similar order of magnitude. This will be   shown  in Section~\ref{numerical}  were we study dependence of the hybrid spectrum on  the mean field parameter, $\omega$ ({\it cf.} Eq.~(\ref{vac})).  We also restrict our study to lowest multiplets,  since bare bound state spectrum 
   is expected to be renormalized above open decay channels. 
   
   The two lowest $(L_q,J_g)$ multiples , $(L_q,J_q) = (0,1), (1,1)$ lead to the following set of states. 
    For $J_g=1$ there two possible gluelumps, the ground state with $J_g^{PC} = 1^{+-}$ and gluon in the 
     TE mode and the first excited gluelump state with $J_g^{PC}=1^{--}$ and the TM  gluon. 
 Coupling the TE gluon with the color octet $Q{\bar Q}$ state in $L_q=0$, $S$-wave orbital leads to a hybrid state with the  intermediate  angular momentum ${\bf j} = {\bf L}_q  + {\bf J}_q= 1$. After adding the quark spin $S_q=0,1$,  in absence of  spin-dependent interaction and weak mixing between different quark orbitals,  we obtain four low lying hybrids with quantum numbers, $J^{PC} = 1^{--}$ for  $S_q=0$ and $J^{PC} = 0^{-+}, 1^{-+}, 2^{-+}$ for $S_q=1$.  It is worth noting
   that hybrid with exotic quantum numbers $1^{-+}$ appears in this lowest multiplet and is predicted to have the $Q{\bar Q}$ pair in spin-1.  The TM gluon coupled to  the octet $Q{\bar Q}$ state would lead to four hybrids with $J^{PC} = 1^{+-}$  for   $S_q=0$ and $J^{PC} = 0^{++},1^{++},2^{++}$ for $S_q=1$. The $L_q=1$ ${Q\bar Q}$ multiplet, is expected to be at a similar energy if orbital excitations of the quarks is energetically comparable to orbital excitation of the quasi-gluon.  

\subsection{ Numerical Results} 
\label{numerical}
Solving the radial Shr\"odinger equation, ~(\ref{seq}), even for fixed $L_q,J_g$ is quite challenging numerically.   
  We proceed with the following numerical approximation. First we use a single harmonic oscillator  quark wave function  to solve for the low-lying 
   gluon energy in the background of either $S$ ($L_q=0$) or $P$ ($L_q=1$) wave quarks. The size of the quark wave function is determined by minimizing the total hybrid energy. In the next step and to validate single harmonic oscillator approximation for the  quark wave function , we use the solution for the gluon wave function from the first step and solve 
    for  the quark wave function.  We find variation in the hybrid energies to be less then a percent. 
 Taking advantage of the gaussian quark wave function, we can introduce an effective potential for the gluon field 
  and, as discussed earlier make  the connection between  hybrid and gluelump states. In this approximation 
   hybrid  can be viewed as a single quasi-gluon moving in an effective potential originating 
    from the slow moving quarks.    In the Appendix we give explicit formulas for  the  gluon interactions 
     involving  $S$-wave quarks.  
    \begin{figure}
\begin{center}
\includegraphics[width=2.8in,angle=270]{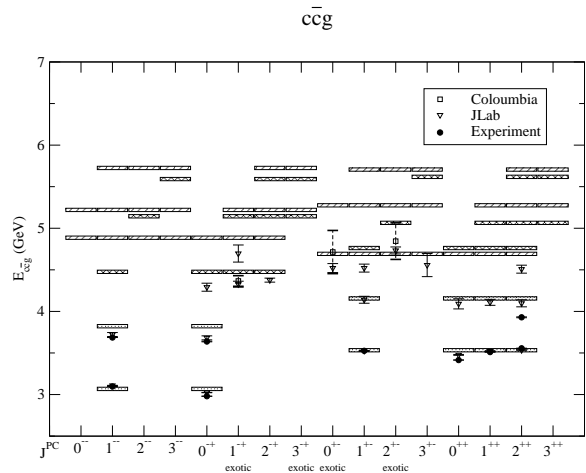}
\caption{\label{cc} Charmonium (solid boxes) and charmonium hybrid spectrum compared with data (where available) or lattice computations. Single dashed boxes are the $c{\bar c}g$ hybrids dominated by the $P$-wave quarks, all other have the $Q{\bar Q}$ pair in the relative $S$-wave orbital. } 
\end{center} 
 \end{figure} 
    \begin{figure}
\begin{center}
\includegraphics[width=2.8in,angle=270]{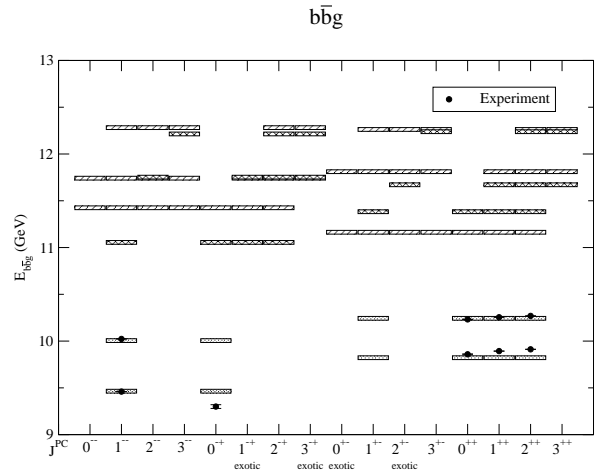}
\caption{\label{bb}  Same as Fig.~\ref{cc} for bottomonium. } 
\end{center} 
 \end{figure}    
 In Fig.~\ref{cc} full boxes show our results for the spin-averaged masses of ordinary quarkonia. As discussed in Sec.~\ref{g-h} the $0^{-+}$ and $1^{--}$ states were used to fix the quark quark mass.  We find the first  radial $S$-wave excitation  at $M_{0'^{-+}} = 3.82\mbox{ GeV}$ and the center of mass of the $P$-waves at $M_{1^{+-}} = M_{0^{++}} = M_{1^{++}} = M_{2^{++}} = 3.53\mbox{ GeV}$,  which compare favorably to the experimental values as seen from Table~\ref{table-sqq}.
 
 \begin{table*}[htdp]
\caption{$Q{\bar Q}$ charmonium spectrum}
\begin{center}
\begin{tabular}{|c|c|c|c|c|c|} \hline
$J^{PC}$&  Exp. [GeV] & This work [GeV] & $J^{PC}$& Exp [GeV]  & This work  [GeV] \\  \hline
$0^{-+}$ & 2.980(1) & 3.07 &$0'^{-+}$& 3.637(4) &  3.82  \\ \hline
$1^{--}$ &  3.097(0)&3.07&$1'^{--}$& 3.686(0)& 3.82 \\ \hline
$1^{+-}$ & 3.526(0)  & 3.53&$1'^{+-}$ & $-$& 4.16 \\ \hline
$0^{++}$ & 3.415(0)& 3.53 &$0'^{++}$& $-$& 4.16 \\ \hline
$1^{++}$ & 3.511(0)  & 3.53 &$1'^{++}$& $-$ & 4.16 \\ \hline
$2^{++}$ & 3.556(0)  & 3.53 &$2'^{++}$& 3.929(5) & 4.16 \\ \hline
\end{tabular}
\label{table-sqq} 
\end{center}
\end{table*}

\color{black}
 As discussed in ~\cite{Guo:2007sm} we determine gluon mean field parameter $\omega(k)$ which appears in the ansatz for the vacuum, Eq.~(\ref{vac}) by fitting the gluelump spectrum. Using $\omega(k)$ specified by 
 model-3 in ~\cite{Guo:2007sm} for the masses of lowest two gluelump states $J_g^{PC} = 1^{+-}$ and $1^{--}$ , gives 
   $0.89\mbox{ GeV}$ and $1.29\mbox{ GeV}$ compared to 
lattice  results of  $0.87(15)\mbox{ Gev}$ and  $1.25(16)\mbox{ Gev}$, respectively. In comparison, the original vacuum model of Ref.~\cite{Szczepaniak:2001rg,Szczepaniak:2003ve}, referred  to in ~\cite{Guo:2007sm} as model-1 results in higher gluelump masses of  $2.05\mbox{ GeV}$ and $2.19\mbox{  GeV}$, respectively. 

    \begin{figure}
\begin{center}
\includegraphics[width=3in,angle=0]{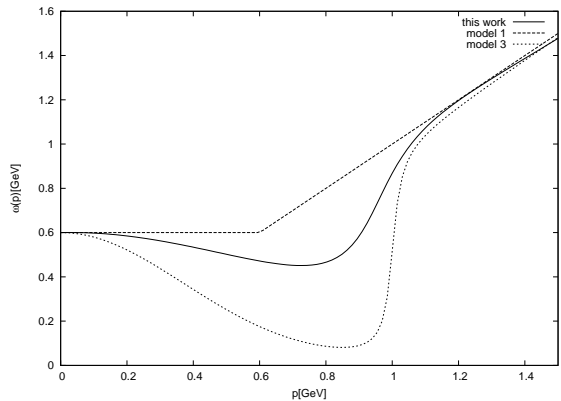}
\caption{\label{wg} Vacuum wave functional parameter $\omega(k)$, model-1 and model-3 were discussed in ~\cite{Guo:2007sm}. The dotted line is used here and give the best fit to the lowest gluelump and $1^{-+}$ hybrid. } 
\end{center} 
 \end{figure} 
  We can use the gluelump and ordinary $Q{\bar Q}$ masses  to obtain a simple estimate 
 for the  hybrid  spectrum.  For hybrids with the TE gluon {\it i.e.}  built on top of the ground state, $1^{+-}$ gluelump and the  $S$-wave $Q{\bar Q}$ state we expect the exotic $1^{-+}$ charmonium hybrid at   $M_{1^{-+}} \sim 
 m_g +  2m_{Q} + m_{Q} \langle v^{2} \rangle  \sim   0.87(2.05) +    2.3 + 0.5 = 3.67 (4.85)  \mbox{ GeV}$ for  model-3 (model-1) were we used  $\langle v^{2} \rangle  \sim 0.4 \sim 0.5$.    The recent lattice calculation of ~\cite{Dudek:2007wv}  palce the $1^{-+}$ charmonium exotic at  $4.33\mbox{ GeV}$, {\it i.e.} between our  two models of the vacuum wave functional. Thus finally  we vary $\omega(k)$  to  obtain   masses closest to both the ground state gluelump and the $1^{-+}$  hybrid. The corresponding  $\omega$ is shown in Fig.~\ref{wg} for which we find the $1^{-+}$ hybrid 
  at $4.47\mbox{ GeV}$ and the two lowest gluelumps at $M_{1^{+-}}  = 1.72 Gev$ and $M_{1^{--}}  = 2.12 Gev$.

\begin{table*}[htdp]
\caption{Charmonium hybrids with $S$-wave $Q{\bar Q}$ pair compared with lattice results were avilable. 
The second column shows masses all states in a given multiplet (third  column)  from   the Coulomb gauge calculation. In the forth column we give the lattice results that are closest to our predictions. The states marked by $[?]$ (if exist) have not been resolved on the lattice. Lattice computations find lower mass states for these quantum numbers. it is possible that these  could be interpreted as $Q{\bar Q}$ states. The exception is the $3^{-+}$ for which no lattice state was found. }
\begin{center}
\begin{tabular}{|c|c|c|c|c|} \hline
$J_{g}^{P_{g}}$&   This work [GeV] & $J^{PC}$& Lattice~\cite{Dudek:2007wv} [GeV] \\  \hline
$1^{+}$ &  4.476 &$0^{-+},1^{-+},2^{-+},[1^{--}]$& 4.291(48),4.327(36),4.376(24), [?] \\ \hline
$1^{-}$ & 4.762 &$1^{+-},2^{++},[0^{++},1^{++}]$ & 4.521(48),4.508(48), [?,?] \\ \hline
$2^{+}$ &  5.144 &$1^{-+},[2^{--},2^{-+},3^{-+}]$ & 4.696(103), [?,?,?]\\ \hline
$2^{-}$ &  5.065 &$2^{+-},[1^{++},2^{++},3^{++}]$& 4.733(42), [?,?,?] \\ \hline
\end{tabular}
\label{table-s}
\end{center}
\end{table*}

\begin{table*}[htdp]
\caption{Same as in Table~\ref{table-s} for $Q{\bar Q}$ in the $P$-wave }
\begin{center}
\begin{tabular}{|c|c|c|c|c|} \hline
$J_{g}^{P_{g}}$&   This work [Gev] & $J^{PC}$&  Lattice~\cite{Dudek:2007wv} [Gev] \\  \hline
$1^{-}$ &  4.886 &$0^{-+},1^{-+},2^{-+},$ [7 more]& 4.291(48),4.327(36),4.376(24), [?]  \\ \hline
$1^{+}$ & 4.692 &$0^{+-},1^{+-},2^{++},$ [7 more] & 4.521(54),4.521(48),4.508(48), [?]   \\ \hline
$2^{-}$ &5.221 &$1^{-+},$ [11 more] & 4.696(103), [?] \\ \hline
$2^{+}$ & 5.276 &$2^{+-},$ [11 more]& 4.733(42), [?] \\ \hline
\end{tabular}
\label{table-p}
\end{center}
\end{table*}

The final results for the charmonium and bottomonium spectrum are shown in Fig.~\ref{cc} and Fig.~\ref{bb} respectively. As discussed earlier, if mixing between $(L_q,J_g)$ multiplets is excluded and with hyperfine interactions turned off we find  $4$ $Q{\bar Q}g$ degenerate states for $L_g=0$ and 12 degenerate states for $L_g=1$. We have computed the effects of  $O(1/m)$ interactions which include spin dependent terms and find them to contribute at the level of $30-50\mbox{ MeV}$ to the charmonium states which is certainly below the  accuracy of the variations in the mean field wave functional.  Our results are compared to experimental data and/or lattice calculations where available. The particular degeneracy pattern we find seems to suggest that the $PC=-+$ triplet  of states found in lattice simulations at $4.3\mbox{ GeV}$ does indeed contain the $1^{-+}$ exotic~\cite{Dudek:2007nj}. 
In Tables~\ref{table-s},\ref{table-p} we give our predictions for the  masses  of the $c{\bar c}$g hybrids and the possible identification with the lattice states from ~\cite{Dudek:2007wv}.

 \begin{figure}
\begin{center}
\includegraphics[width=2.8in,angle=270]{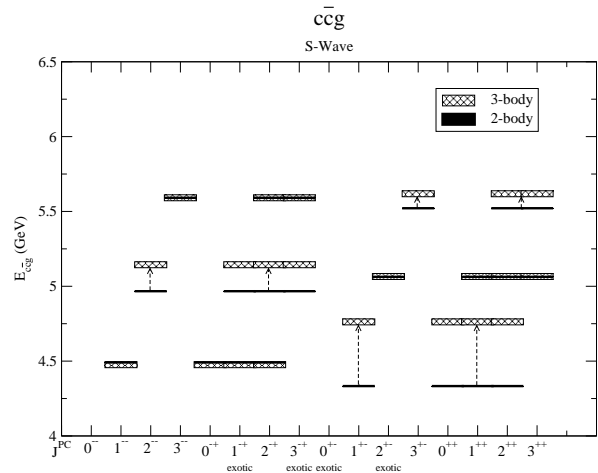}
\caption{\label{2-vs-3} The effect of the irreducible three body potential on the charmonium hybrid spectrum with $S$-wave quarks.  Sold lines represent the spectrum form  Eq.~(\ref{seq}) with $V^3$ in Eq.~(\ref{vq}) removed. } 
\end{center} 
 \end{figure} 
Finally in Fig.~\ref{2-vs-3} we show the effect of the irreducible  three body interaction, $V^3_{QQg}$, that is germane  to the Coulomb gauge. It is this interaction that is responsible for producing the inverted parity ordering of the gluelump spectra,{\it i.e} producing the $1^{+-}$ gluelump below the $1^{--}$. It leads to analogous parity-inversion in the $Q{\bar Q}g$ quarkonium spectra, {\it i.e.} it leads to the  $1^{--}, 0^{-+}, 1^{-+}, 2^{-+}$ multiplet below $1^{+-}, 0^{++}, 1^{++}, 2^{++}$. 

\section{ Summary and Outlook} 
\label{outlook}  
We  studied spectrum of heavy quarkonium hybrids in Coulomb gauge QCD with gluon degrees of freedom in the mean field approximation. In general, we have found a reasonable agreement with lattice data.  In particular we find the 
 $1^{-+}$ exotic charmonium state at $4.47\mbox{ GeV}$. Our predictions seem to be systematically higher compared to lattice which is most likely a reflection of a deficiency of the gaussian, mean filed, vacuum wave functional. This needs to be improved in particular by the effects of  vortex field configurations, wildly believed to be revenant for confinement. 
  The framework however seem to be well suited for further investigations of  quarkonium and hybrid structure, including  mixing and transitions.

\section{Acknowledgment}
We would like to thank  J.~Dudek for discussions of the lattice results and H.~Reinhardt  for continuing discussions of the Coulomb gauge QCD. This work was supported in part by the US Department of Energy grant under  contract DE-FG0287ER40365. APS also thanks the  I.N.F.N and University of Genova for hospitality  during preparation of this work.

\appendix

\section{Hamiltonian Matrix Elements} 
The matrix elements of $V^2_{Q{\bar Q}g}$ and $V^3_{Q{\bar Q}g}$ in Eq.~(\ref{vq}) are computed using gaussian wave functionals  for the $Q{\bar Q}$ radial wave function. 
\begin{eqnarray} 
&& V^{2,3}_{Q{\bar Q}g}(k,k')  \nonumber \\
& = & \int \frac{q^2 dq}{(2\pi)^3}  \frac{q'^2 dq'}{(2\pi)^3} \Psi^{L_q}_Q(q) V^{2,3}_{Q{\bar Q}g}(k,q,\alpha;k',q',\alpha') \Psi^{L_q}_Q(q' ) \nonumber \\
\end{eqnarray}
In  particular for the $S$-wave, given by Eq.~(\ref{s-wave}) we find   
\begin{widetext}
\begin{eqnarray}
V^{2}_{Q \bar{Q} g} (k,k') &=&\delta_{J'_g, J_g}    N_c
\left(\sqrt{\frac{\omega(k)}{\omega(k')}}+\sqrt{\frac{\omega(k')}{\omega(k)}}\right) \sum_{l}  \frac{1}{2}[1+P(-1)^{l}] \langle J_g 1;1-1 |l 0 \rangle^2
\nonumber \\
&\times&  (2\pi) \int_{-1}^1 d \hat\k\cdot\hat\k'  P_{l}(\hat\k\cdot\hat\k') \widetilde{V}_{CL}
(|\mathbf{k} -\mathbf{k}'|) e^{-\frac{ |\mathbf{k} -\mathbf{k}'|^2}{16
\alpha_{Q}^2}}.
\end{eqnarray}
and  
\begin{eqnarray}
 V^{3}_{Q{\bar Q}g}  (k,k') &=& -  \frac{N_c^2}{8}
 \int d^2 \widehat{\mathbf{k}}
d^2 \widehat{\mathbf{k'}} \frac{1}{\sqrt{\omega(k) \omega(k')}}
\sum_{ m_{1},m_{2},\sigma, \sigma'} \chi^{\ast
JMPC}_{m_{1},m_{2},\sigma} (\hat{\mathbf{k}},\alpha)
\chi^{J'M'P'C'}_{m_{1},m_{2},\sigma'}
(\hat{\mathbf{k}'},\alpha')  \nonumber \\
&\times&  ( 4
  \pi) \int
 \frac{d^3 p}{(2 \pi)^3}  K^{(2)} (|\mathbf{k'}+\mathbf{p}|,|\mathbf{k}+
\mathbf{p}|,| \mathbf{p}|)  [ 2  e^{-\frac{|2 \mathbf{p} +
\mathbf{k} +\mathbf{k'}|^2}{ 16 \alpha_{Q}^2}}+e^{-\frac{|\mathbf{k}
-\mathbf{k'}|^2}{ 16 \alpha_{Q}^2}}] \mathbf{p} \cdot \epsilon_{-k',
\sigma'} \mathbf{p} \cdot \epsilon^{*}_{-k, \sigma} .\end{eqnarray}
\end{widetext}
where ~\cite{Guo:2007sm},  $  K^{(2)} (k,p,q) = \widetilde{V}_{CL} (k)  \widetilde{V}_{C}(p) \widetilde{V}_{C}(q)  +\widetilde{V}_{C} (k)  \widetilde{V}_{CL}(p) \widetilde{V}_{C}(q) + \widetilde{V}_{C} (k)  \widetilde{V}_{C}(p) \widetilde{V}_{CL}(q) $. With hybrid parity and charge conjugation satisfying $P=(-1)^l$ {\it i.e.} the possible values of $l$ are constrained by parity of a state, $C = (-1)^{S+1}$  where  is $S=0,1$ is the spin of the $Q{\bar Q}$ pair.  
To order $O(1/m)$ the other two interactions in Eq.~(\ref{h1}) give 
\begin{widetext}
\begin{eqnarray}
V_{A^2} ^{S}(k,k') & = & \frac{1}{4 m } (C_F-\frac{1}{2 N_c})
\frac{1}{\sqrt{ \omega_k  \omega_{k'}}}
 \sum_{l}  [1+P (-1)^{l}]
    \langle J_g 1;1-1 |l 0 \rangle ^2  \nonumber \\
    &\times&    \int d \hat\k\cdot\hat \k' P_{l}(\hat\k\cdot\hat \k' )  (4 \pi)^2 \alpha(|\mathbf{ k}-\mathbf{ k}'|)  e^{-\frac{ |\mathbf{ k}-\mathbf{ k}'|^{2} }{16 \alpha_Q^2}}  \nonumber  .\nonumber
\end{eqnarray}
\end{widetext} 
for spin-idependent contact interaction and, 
\begin{widetext} 
\begin{eqnarray}
V_{\sigma B}^S(k,k') &=& \frac{3 N_c}{4 m} \frac{1}{\sqrt{\omega_k
\omega_{k'}}}
\sum_{l}  [1+P (-1)^{l}] (-1)^{1+l +J} (2 S+1)(2 l+1) \langle 1 1; l 0 |J_g 1\rangle^2 \nonumber \\
   &\times&  \left \{
\begin{array}{ccc} S & \frac{1}{2} & \frac{1}{2} \\ \frac{1}{2} & S& 1 \end{array} \right
\}   \left \{
\begin{array}{ccc} 1 & J_g & J_g \\ l & 1& 1 \end{array} \right
\}  \left \{
\begin{array}{ccc} 1 & J_g & J_g \\ J & S& S \end{array} \right
\}
 \int d \hat\k\cdot\hat \k'   P_{l}(\hat\k\cdot\hat \k' ) (4 \pi)^2   \alpha(|\mathbf{ k}-\mathbf{ k}'|)
 e^{-\frac{|\mathbf{ k}-\mathbf{ k}'|^{2}}{16 \alpha_Q^2}} 
   \nonumber
\end{eqnarray}
\end{widetext}

for spin-dependent contact interaction. Here 
\begin{equation}
\alpha(p) = \frac{4 \pi Z}{\beta^{{\frac{3}{2}}}   \ln^{\frac{3}{2}} (\frac{p^{2}}{\Lambda^{2}_{QCD}}+c) }
\end{equation}
with $Z=5.94, c = 40.68,  \Lambda_{QCD}=0.25 \mbox{ GeV}$ arising from the fit to  the Coulomb part of the QCD 
 potential~\cite{Guo:2007sm}.


\end{document}